\documentclass[12pt,a4paper]{article}
\usepackage{amsmath,amssymb, amsfonts, amsthm}

\title{Quantum probabilities and violation of CHSH-inequality 
from classical random signals and threshold type properly calibrated detectors}

\author{Andrei Khrennikov\\
International Center for Mathematical Modelling
\\in Physics and Cognitive Sciences\\t
Linnaeus University,  V\"axj\"o, S-35195, Sweden}

\begin{document}

\maketitle

\begin{abstract} 
We present a purely wave model (based on classical random field) which reproduces quantum probabilities (given by the 
fundamental law of quantum mechanics, Born's rule) including probabilities for joint detection of a pair of 
quantum observables (e.g., spin or polarization projections). The crucial point of our approach is that the presence of  detector's threshold and calibration 
procedure have to be treated not as simply experimental technicalities, but as the basic counteparts of the theoretical model. The presence 
of the background field (vacuum fluctuations) is also  the key-element of our prequantum  model. It
is of the classical signal type and the methods of classical signal theory (including statistical radiophysics) are used for its development. We stress that our prequantum model is not objective, i.e., the values of observables (clicks of detectors) cannot be 
assigned in advance, i.e., before measurement. Hence, the dilemma, nonobjectivity or nonlocality, is resolved in favor of nonobjectivity 
(our model is local of the classical field type).
In particular, we reproduce the probabilities for the EPR-experiment for photon polarization and, hence, violate CHSH inequality 
for classical random signals (measured by the threshold type and properly calibrated detectors acting in the presence of the background field).
\end{abstract}

\section{Introduction}

The Bell's inequality \cite{B} plays a crucial role in modern quantum mechanics and, especially, quantum information (QI).
Its violation has not only theoretical consequences (nonlocality, nonobjectivity of quantum observables), but also applications,
e.g., to quantum cryptography \cite{GREG}. Its violation was experimentally confirmed \cite{AS}, \cite{Z} (although
 there are still loopholes, see, e.g., \cite{GREG}, \cite{AD}--\cite{CONT} for discussions). There are no doubts in results of experiments. However, a proper interpretation
of these results is till the subject of intensive debates, see, e.g., \cite{CONT}, \cite{PR}, \cite{PLOT}. By the commonly accepted interpretation it was proved 
that quantum observables are either nonlocal or/and nonobjective. Although the Bell's test does not provide a possibility to distinguish
nonlocality from nononbjectivity, the majority of QI-people made (intuitively) their choice in favour of 
{\it nonlocality.} This viewpoint has been criticized by some authors, e.g.,  \cite{Fine}--\cite{Raedt}, see also \cite{CONT} for extended bibliography.  The majority of authors criticizing the conventional interpretation of violation of Bell's inequality tried to save {\it both locality and objectivity} (a possibility to assign to a system the values of physical observables before measurement). 
This is not my approach.  I agree (although this contradict to my own ``old papers''\cite{OLD}) that it is impossible to combine locality and realism 
and reproduce quantum probabilities for entangled systems; in particular, to violate Bell's type inequalities, e.g., the CHSH inequality.
In this paper I present a local, but {\it nonobjective} classical model violating the Bell's type inequality for
{\it probabilities of joint detections}, namely, CHSH-inequality.

 Nonobjectivity of observables is typically considered as an intrinsically quantum feature. Bohr emphasized the role of measurement context in quantum measurements. At the same time classical physics is often associated with 
one special model, {\it classical statistical mechanics,} which is definitely objective. It is forgotten that, besides  classical statistical mechanics, there exists another important classical model -- {\it classical field theory.} In this paper we show that the usage 
of the threshold type detectors operating with (classical) random signals makes observables for classical signals nonobjective. 
Hence, Bohr was right, the experimental context plays a crucial role in QM. However, it also plays a similar role in some classical 
models of the wave-type. We call our model {\it threshold signal detection model}, TSD.

TSD definitely has important  consequences for quantum foundations: QM can be treated as a part 
of classical signal theory. Hence, opposite to Bohr's claim, QM may be incomplete; opposite to Bell's claim, it may be local;
opposite to Einstein's claim, it need not be objective. Of course, in physics the creation of a theoretical model, in our case TSD,  is not the end of the story. The final word always should be said by experimenters. To confirm TSD experimentally, experimenters have to be able to measure  components of (classical) fields corresponding to quantum particles at so to say ``prequantum level'' (for example,
electric and magnetic components of the photon). 
 
The impact of TSD to QI is a more complicated problem. Since QM can be embedded in classical signal
theory, it seems that QI can be considered as a part of classical information theory. Surprisingly this is not the case.
QI was elaborated to operate with incomplete information\footnote{This is the interpretation of QM and QI based on TSD. It differs
crucially from the orthodox Copenhagen interpretation.} provided by quantum observables. Its operations and consequences cannot
be directly derived from classical signal theory.  Nevertheless, it is clear that after creation of  TSD the Bell's test
cannot be considered guarantying   100\% security of the basic quantum cryptographic protocols.

We list the basic assumptions of TSD: 

\bigskip

(a) prequantum signals have a special temporal structure of correlations
given by (\ref{BE0dja})--(\ref{BE0dj3a});

\medskip

(b) detectors are of the threshold type;

\medskip

(c) detectors are properly calibrated to eliminate the contribution of the random background field;

\medskip

(d) instances of clicks of detectors for measurements on correlated signals match each other;

\medskip

(e) stochastic processes inducing quantum probabilities and correlations are Gaussian.

\medskip

Thus the temporal structure plays an important role in our treatment of Bell's inequality, cf. \cite{Hess}, \cite{Raedt}, \cite{CONT} .

The usage of the threshold type detectors ruins objectivity of quantum observables. It is possible to determine
only instances of detectors' clicks; in TSD we are not able to represent quantum observables in the Bell's form\cite{B}:
\begin{equation}
\label{FR}
a=a(\lambda),
\end{equation} 
where $\lambda$ is so called hidden variable. 

The calibration of detectors is not a technicality.
This is a basic element of TSD; quantum correlations are obtained through discarding the contribution of the
random background field. This field is fundamental and it is impossible to distil it from the quantum signal (quantum
system). We are only able to eliminate it through the measurement procedure, via calibration. 

It is well known 
that in real EPR-Bohm experiments clicks of detectors for  channels corresponding to entangled photons have to match each
other. In practice, this is done with the aid of time window.\footnote{A possibility to violate Bell's inequality for
a classical corpuscular model by using  the time window was explored in \cite{Raedt}.} Typically this matching is considered as
an experimental technicality. However, as it was shown in \cite{KHR}, this is a foundational question related to 
the projection postulate in QM (the difference between L\"uders postulate and the original von Neumann postulates for 
measurements on composite quantum systems). In TSD the condition (d) is also fundamental.

TSD can be considered as measurement theory for recently developed {\it prequantum classical statistical field theory}, PCSFT, \cite{KH1}. 
The latter reproduced all quantum averages and correlations including correlations for entangled quantum states. In particular, 
PCSFT correlations violated Bell's inequality. 

{\it The message of PCSFT in a nutshell is that (i) quantum systems may be mapped on classical stochastic systems even if they are capable of nontrivial quantum manifestations, and that (ii) this shows that the aforesaid phenomena should be regarded more classical than it is commonly believed.} 

Examples of mappings with the stated properties are well known: the $Q$-representation for linear bosonic systems, and the so-called positive-$P$ representation for nonlinear ones. The $Q$-function of an electromagnetic field in a quantum state is positive, which does not preclude such field from showing violations of Bell inequalities in A. Aspect's experiment.
The main problem for matching of PCSFT and conventional QM was that PCSFT (nor other aforementioned models) was not able to describe
probabilities of discrete clicks of detectors. In particular, PCSFT  is  theory of correlations of {\it continuous signals}.
``Prequantum observables''  are given by quadratic forms of signals. These forms are unbounded and this is not surprising that correlations of such observables can violate Bell's type inequalities, see \cite{CONT} for discussion and an elementary example.
(The condition of coincidence of ranges of values of quantum observables and corresponding ``prequantum variables''
plays a crucial role in Bell's argument.)
TSD solved the measurement problem of PCSFT. In the same way as in Bell's consideration, TSD operates with discrete observables. In particular, in the case of photon polarization (its projection to a fixed axis) TSD operates with dichotomous variables taking values
$\pm 1.$ 

\section{Resolution of dilemma: nonlocality or nonobjectivity?} 

This section is devoted to the general discussion on Bell's inequality, nonlocality, nonobjectivity, contextuality. As was emphasized
in introduction, our classical field type model, PCSFT, endowed with the corresponding measurement model TSD is a local, but nonobjective. Hence, the  dilemma ``nonlocality or nonobjectivity?'' is resolved in favor of nonobjectivity. (In our framework 
one cannot use the functional representation of quantum observables (\ref{FR}) and, hence, it is not surprising that Bell's inequality 
can be violated.) In this section we couple
this (yet purely theoretical) prediction with experimental studies in quantum foundations, namely, experiments on quantum contextuality
\cite{CONT1}, \cite{CONT2}. Although these experiments have no direct relation to PCSFT/TSD, their results might be interepreted as supporting nonobjective
``prequantum'' models. 

We state again the basic assumptions of Bell's argument:

\medskip
 
(R) {\bf Realism:} A possibility to assign to a quantum system the values of observables before measurement.

 From the philosophical viewpoint this is not precisely the definition of realism (objectivity). To be real (objective), it is enough to exist, without any relation with experiment. Such ``ontic realism'' is formalized through the principle of value definiteness:

(VD) {\it All observables defined for a QM system have definite values at all times.}

 However, Bell used  ``measurement realism'' which 
we presented in (R). If the values of physical observables were existing, but not coinciding with results of measurement, then Bell's 
consideration would not imply Bell's inequality, see \cite{CONT} for analysis and examples. In philosophic literature (R) is often  referred as a {\it principle of faithful measurement} (FM) \cite{EE}.

\medskip

(L) {\bf Locality:} No action at the distance.

\medskip

Therefore every one (who accepts that experiments are strong signs that local realism has to be rejected)
has to make the choice between:

\bigskip

(NONL) Realism, but nonlocality (the original Bell's  position).

\medskip

(NR) No realism (nonobjectivity) and locality  (the original Bohr's position).

\medskip

(NONL+NR) Nonlocality + nonobjectivity.

\medskip

The last possibility, (NONL+NR),  seems to be too complex to happen in nature. Of course, one cannot completely 
reject that nature is so exotic. However, to resolve all problems one need not make this assumption, either
nonlocality or nonobjectivity is enough. The (NONL+NR)-interpretation of experimental results is definitely
non-minimalistic and it can be rejected, e.g., by the [Occam's razor]-reason.

Hence, one has to make his choice: either nonlocality or nonobjectivity;
either De Broglie-Bohm-Bell or Bohr-Heisenberg-Pauli position. We  state again that the Copenhagen interpretation of 
quantum mechanics had nothing to do with nonlocality. Bohr advertised the position that the values of quantum 
observables are ``created'' in the process of interaction of quantum systems with measurement devices. Hence, 
the main point was nonobjectivity.

It is typically assumed that the present experimental situation does not provide us a possibility to make 
the choice. And this is correct if one explores only experiments of the EPR-Bohm type in which realism and 
locality are mixed. 

However,  recently exciting experiments testing {\it quantum contextuality}  were performed \cite{CONT1}, \cite{CONT2}: they supported 
the thesis that quantum mechanics is contextual. 

We point that  {\it contextuality implies nonobjectivity!}\footnote{We state again that we understood objectivity (realism)  
as ``measurement objectivity'' (realism) -- the discussion  after the definition of (R). Contextuality does not imply the violation of 
the principle of value definiteness (VD).}
 In the contextual situation it is impossible to assign values
of physical observables before measurement. Therefore the experiments \cite{CONT1}, \cite{CONT2} can be considered as supporting 
nonobjectivity. This experiment is about nonojectivity of results of measurements for a single particle.

Now I present the following considerations which seem to be logically justified. If already a single particle exhibits
lack of objectivity, then it is reasonable to assume that the situation cannot be improved by consideration of a pair of particles. Hence, it is reasonable to assume nonobjectivity in the EPR-Bohm experiment. 
This implies that among two alternatives, (NONL) or (NR), the latter is essentially more justified than the former.

We can summarize the arguments presented in this session:

\bigskip

 {\it Recent experiments on quantum foundations can be considered as supporting the original Bohr's position 
-- quantum observables are nonobjective, their values cannot be assigned before measurement.
The assumption of nonlocality has to be rejected, since there are no direct experimental evidences of nonlocality
(similar to the tests of nonobjectivity performed in \cite{CONT1}, \cite{CONT2}) and since in the EPR-Bohm experiment it is unnecessary -- under 
the assumption of nonobjectivity.}

\section{From time-correlations in prequantum random signals to quantum probabilities}
\label{S1}

\subsection{The scheme of threshold detection}

Let us consider a complex valued stochastic process (random signal) $\phi(s)= \phi(s, \omega)$ with zero average, 
$E\phi(s)=0$ for any $s.$ 
The quantity 
\begin{equation}
\label{BE0}
{\cal E}(s, \omega)= \vert  \phi(s, \omega)\vert^2 
\end{equation}
is the signal energy at the instant of time $s.$ If signals corresponding to quantum systems were smooth enough, then the detection procedure under consideration would be reduced to 
the condition of the energy level approaching the detection threshold, say ${\cal E}_d >0.$ The  isntant of time $\tau$ corresponding to 
the signal detection (``click'') is determined by the condition:
\begin{equation}
\label{BE0a}
{\cal E}(\tau, \omega)=  {\cal E}_d.
\end{equation}
We remark that the instant of the signal detection is a random variable:
$$
\tau=\tau(\omega).
$$
Mathematically our aim is to find average of the instance of detection, $\bar{\tau}= E \tau.$ 
The quantity $1/\bar{\tau}$ will be used to find the probability of detection, ``how often the detector produces clicks,'' see section \ref{S2}.

However, classical random signals corresponding to quantum states are very singular (because  of the contribution of the background field of the
white noise type) and the value of a signal at the finxed instance of time is not defined (at least we cannot be sure that it is defined for allmost 
all $\omega).$ Therefore, instead of the signal's energy value at the fixed instance of time (\ref{BE0}), we shall use  the analog of the threshold approaching condition 
for a properly smoothed signal. We shall consider smoothing in the $L_2$-space. This smoothing matches the real detection procedure.
In reality, a detector cannot determine the  signal's energy at the fixed instance of time. Any detector is based on the integration of signals.

Suppose that the detection procedure  is based on the {\it integration window} given by the step-function
%\begin{displaymath}
\begin{equation}
\label{KAPPA}
g(s) \equiv g^{\kappa}(s) = \left\{ \begin{array}{cl}
1/\sqrt{\kappa}, & s\in  [0, \kappa]  \\
0 \; \; \; \; , & s \not\in  [0, \kappa]
\end{array}\right. 
\end{equation}
%\end{displaymath}
where $\kappa> 0$ is a small parameter (of the detector).
We remark that $\Vert g \Vert=1$ (the $L_2$-norm).

Mathematically the detection procedure is described in the following way. Consider the $\kappa$-smoothed signal
\begin{equation}
\label{BEb0}
\phi^{\kappa}(u, \omega)= \int_{-\infty}^{+\infty}  \phi(s, \omega) g(u - s) ds= 
\frac{1}{\sqrt{\kappa}} \int_{u -\kappa}^u  \phi(s, \omega) ds
\end{equation}
and its energy
\begin{equation}
\label{BEb}
{\cal E}(u, \omega; \kappa)= \vert \phi^{\kappa}(u, \omega) \vert^2.
\end{equation}
Now the instant of time of signal's detection  $\tau$ 
is determined by the condition, cf. (\ref{BE0a}):
\begin{equation}
\label{BE0c}
{\cal E}(\tau, \omega;   \kappa)=  {\cal E}_d.
\end{equation}

We consider the special class of random signals having zero averages ($E\phi(s)=0$ for any $s).$   Suppose that the covariance function of $\phi(s)$ has the following form:
\begin{equation}
\label{BE0d}
E \phi(s_1) \overline{\phi(s_2)}= \sigma^2 \delta(s_1-s_2) \sqrt{\vert s_1 s_2\vert}. 
\end{equation}
(The role of the parameter $\sigma^2$ will become clear in section \ref{S2}, Remark 3.).
We find the average of the energy ${\cal E}(\tau, \omega;  \kappa)$ of this signal. 
We have 
$$
E{\cal E}(\tau, \omega;  \kappa)=
$$
$$
\frac{1}{\kappa} E \Big\vert  \int_{\tau -\kappa}^\tau ds \phi(s, \omega) \Big\vert^2 =
\frac{1}{\kappa}  \int_{\tau -\kappa}^\tau \int_{\tau -\kappa}^\tau    E \phi(s_1, \omega) \overline{\phi(s_2, \omega)} ds_1 ds_2=
 $$
 $$
 \frac{\sigma^2}{\kappa}  \int_{\tau -\kappa}^\tau \int_{\tau -\kappa}^\tau  \delta(s_1- s_2) \sqrt{\vert s_1 s_2\vert} ds_1 ds_2  =
 \frac{\sigma^2}{\kappa}  \int_{\tau -\kappa}^\tau s ds= \sigma^2(\tau + \kappa).
 $$
We shall use this quantity a bit later. Now we proceed to calculation of the average detection time $\bar{\tau}.$ We take the average
of the equality (\ref{BE0c}) and obtain
\begin{equation}
\label{BE0c1} 
E {\cal E}(\tau(\omega), \omega;   \kappa)=  {\cal E}_d.
\end{equation}
(We recall that the instant of detection $\tau= \tau(\omega)$ is a random variable.)
 To find the quantity in the left-hand side of this equality, we use the formula of total probability
 $$
 E {\cal E}(\tau, \omega;   \kappa)= \int_0^{+\infty}   E [ {\cal E}(\tau(\omega), \omega;   \kappa)\vert \tau(\omega) = \tau]
P(\tau(\omega)  = \tau) d\tau,
$$
where $E[{\cal E} \vert \tau(\omega) = \tau]$ is the conditional expectation of the quantity ${\cal E}$ under the 
condition $\tau(\omega) = \tau.$
The conditional expectation has already been found
\begin{equation}
\label{CE} 
 E [ {\cal E}(\tau(\omega), \omega;   \kappa)\vert \tau(\omega)= \tau] = \sigma^2 (\tau + \kappa).
\end{equation}
Hence, 
$$
E {\cal E}(\tau(\omega), \omega;   \kappa) = \sigma^2 \int_0^{+\infty}  (\tau + \kappa)
P(\tau(\omega)= \tau) d\tau = \sigma^2(\bar{\tau} + \kappa)
$$
\begin{equation}
\label{BE0c1X} 
 = \bar{\tau} \sigma^2 (1+ O(\kappa/\bar{\tau})), \; \kappa/\bar{\tau} \to 0.
\end{equation}
Finally, the averaged condition of detection (\ref{BE0c1}) takes the form:
\begin{equation}
\label{BE0c2} 
\bar{\tau} \sigma^2 (1+ O(\kappa/\bar{\tau})) = {\cal E}_d
\end{equation}
or 
\begin{equation}
\label{BE0c2X} 
 \frac{1}{\bar{\tau}}  \approx  \frac{\sigma^2}{ {\cal E}_d}, \; \kappa/\bar{\tau} \to 0.
\end{equation}

\subsection{Probabilities of clicks in detection channels}
\label{S2}

Hence, during a long period of time $T$ such a detector clicks $N_{\rm{click}}$-times, where
\begin{equation}
\label{BE0c3} 
 N_{\rm{click}} \approx \frac{T}{\bar{\tau}}  \approx  \frac{\sigma^2 T}{ {\cal E}_d}, \; \kappa/\bar{\tau} \to 0.
\end{equation}
To find the probability of detection and match the real detection scheme which is used in quantum experiments \cite{AS} we have to use a proper normalization
of $N_{\rm{click}},$  This is an important point of our considerations. (The normalization problem is typically ignored in standard 
books on quantum foundations, cf., however,   \cite{CONT}.) In QM-experiments probabilities are obtained
through normalization corresponding to the sum of clicks in all detectors involved in the experiment, e.g., spin up and spin down  detectors. 

In QM such a collection of detectors is symbolically represented as quantum observable, say $C.$ In the mathematical formalism observable $C$ is 
represented by the Hermitian operator $\widehat{C}.$  In the case of purely discrete (nondegenerate) 
spectrum, the QM-probabilities of detection are determined by the basis of eigenvectors $\{e_j\}$
of the operator $\widehat{C}$ through the Born's rule:
\begin{equation}
\label{BRULE} 
P_j= \vert \langle \Psi,e_j\rangle \vert^2 
\end{equation}
for quantum systems in the pure state $\Psi$ or more generally, for quantum systems in the mixed state $\rho,$ we have:
\begin{equation}
\label{BRULEX} 
P_j= \rm{Tr} \rho C_j,
\end{equation}
where $\widehat{C}_j$ is the projector onto the vector $e_j,$ 
i.e., $\widehat{C}_j=\vert e_j\rangle \langle e_j\vert.$ In QM the Born's rule (\ref{BRULEX}) is postulated \cite{CONT}.

To reproduce the QM-scheme (in the model in which the spatial degrees of freedom are still absent), 
we conisider a family of stochastic processes $\phi (i,s), i=1,2,...m.$ The signal $\phi$ is split into a family of disjoint channels coupled to detectors $D(i): \phi(s)=(\phi(i,s))_{i=1}^m.$
\footnote{We consider only the detection scheme 
for discrete observables, e.g., spin, $i=+1,$ spin up, and $i=-1,$ spin down.} Thus we have the vector valued random signal $\phi(s).$ 
Suppose that the covariance function of $\phi(s)$ has the following form:
\begin{equation}
\label{BE0dX}
E \phi(i, s_1) \overline{\phi(j, s_2)}= \delta(s_1-s_2) \sqrt{\vert s_1 s_2\vert}  b(ij). 
\end{equation}
(This is simply generalization of (\ref{BE0d}) to the vector valued process,)
Hence, its covariance function can be represented as
\begin{equation}
\label{BE0d1}
B(s_1,s_2) = \delta(s_1-s_2)\sqrt{\vert s_1 s_2\vert} B,
\end{equation}
$B= (b(ij)).$ The matrix $B$ does not depend on temporal correlations; it represents only correlations of internal degrees of freedom
(such as e.g. spin or polarization).
We set 
$$
b(ii)= \sigma_i^2\; \mbox{and} \Sigma^2= \sum_i \sigma_i^2 = \rm{Tr} B.
$$
We repeat the previous detection scheme (based on threshold detectors) for each of this processes, so  
$m$ detectors are involed; the only assumption is that all these detectors have the same detection 
threshold ${\cal E}_b >0.$ We obtain, see (\ref{BE0c3}), 
\begin{equation}
\label{BE0c4} 
 N_{\rm{click}}(i) \approx \frac{T}{\bar{\tau}_i}  \approx  \frac{\sigma_i^2 T}{ {\cal E}_d}, \; \kappa/\bar{\tau_i} \to 0.
\end{equation}
Hence, the total number of clicks:
 \begin{equation}
\label{BE0c5} 
N=\sum_i N_{\rm{click}}(i) \approx  \frac{T \Sigma^2 }{{\cal E}_d} ,
\end{equation}
The probability of detection for the $j$th detector is given by
\begin{equation}
\label{BE0c6} 
P(j)= N_{\rm{click}}(j)/N \approx \frac{\sigma_j^2}{\Sigma^2}.
\end{equation}
In fact, this is the Born's rule. Consider the matrix 
\begin{equation}
\label{BE0d2}
\rho= B/\rm{Tr} B= (b(ij)/\Sigma^2).
\end{equation}
This is the Hermitian positive trace one matrix; so formally it has all properties of the {\it density matrix} used in QM.
In ${\bf C}^n$ take the canonical basis $e_j=(0...1...0);$ set $\widehat{C}_j=\vert e_j\rangle \langle e_j\vert.$ Then the equality for the probability of detection (\ref{BE0c6}) can be written as
\begin{equation}
\label{BE0d3}
P(j) =\rm{Tr} \rho \widehat{C}_j.
\end{equation}
This is the QM-rule for calculation of probabilities of detection.

In the quantum formalism for a given state $\rho,$ density operator, we are able to determine probabilities of detection in corresponding 
channels not only for one fixed observable, the fixed family of disjoint channels, but for any observable, any family of disjoint 
channels. The same feature has our model. We have a stochastic process $\phi(s)$ valued in the $m$-dimensional 
complex Hilbert space $H;$ denote its covariance function by $B(s_1,s_2).$ Suppose that it has the form (\ref{BE0d1})
where $B:H\to H$ is Hermitian positive operator (in general $\rm{Tr} B \not=1).$ This operator describes correlations of internal degrees of freedom in the 
signal $\phi.$ 

Suppose now that {\it all measurement procedures under consideration have the form of projections of the signal $\phi(s)$
onto some orthogonal directions $\{e_j\}$ and the threshold type measurements for components $\phi_j(s)=
\langle\phi(s), e_j\rangle .$} Hence, selection of each measurement of this type is equivalent to decomposition 
of the random signal $\phi(s)$ into orthogonal components.\footnote{In the QM-formalism such a decomposition of a
signal corresponds to the measurement scheme based on the projection postulate. Formally the latter works very well, but 
its origin cannot be explained in physical terms. This brings a bit of mystery to QM-measurement theory: collapse of 
the wave function and so on. In our model the split of a physical signal into a family of signals is the standard operation of the 
classical signal theory, in particular, in classical optics.} Set $b(ij)= \langle e_i\vert B\vert e_i\rangle$ and 
repeat the previous considerations; we obtain (\ref{BE0d3}) for the ``density operator'' $\rho=B/\rm{Tr} B.$
Opposite to the canonical scheme of QM, this operator has a natural interpretation in theory of classical stochastic 
processes (classical signal theory) -- the normalized covariance operator of the internal degrees of freedom of a signal.

\medskip

{\bf Summary.} We considered stochastic processes (with temporal correlations of the special type). They can be used 
to model (classical) random signals with finite-dimesional state space representing non-temporal degrees of freedom, ``internal
degress of freedom.''  The covariance operator for the internal degrees of freedom normalized by its trace can be formally treated
as a density operator, so to say, quantum state. By spliting the random signal into its components corresponding to projections 
onto vectors of an orthogonal basis in the space of internal degrees of freedom we reproduce the detection scheme of QM.

\medskip

{\bf Remark 1.} We stress that the presented derivation was done under the assumption
\begin{equation}
\label{BE011} 
 \kappa/\bar{\tau_i} \to 0.
\end{equation}
Hence, the integration window $\kappa$ has to be essentially smaler than the average time between clicks. This is a natural physical assumption.

\medskip
 
{\bf Remark 2.} We remark that the detection threshold $ {\cal E}_d$ disappeared from the final formula for the probability of detection. However,
the average time between clicks depends linearly on the threshold, see (\ref{BE0c2}).

{\bf Remark 3.} (Dimension analysis) The squared-signal $\vert \phi(s) \vert^2$ has the dimension of energy. 
 From the equality (\ref{BE0d}) we obtain 
that $\sigma^2 \times {\it time} \sim {\it energy}.$ Hence, $\sigma^2  \sim \frac{{\it energy}}{{\it time}} \sim {\it power}.$ The detection threshold 
${\cal E}_d \sim {\it energy}.$ We now comment the equality (\ref{BE0c3}) from the dimensional viewpoint. The number of clicks of a detector, $N_+,$ is 
proportional to signal's power $\sigma^2$ and the duration of the experiment run and inverse proportional to the detection threshold. Hence, {\it signal's 
power (and not its total energy) is crucial for detection.}

\section{Threshold/calibration detection scheme for  classical signals representing entangled quantum systems}
\label{SA} 
The detection scheme presented in this section describes detection of internal degrees of  freedom, e.g., spin components, for pairs of correlated quantum particles. 

Consider a {\it Gaussian}\footnote{We proceed with only Gaussian signals. It may be possible to use non-Gaussian signals.
However, mathematics is essentially more complicated.}  signal with two correlated components (bi-signal) $\phi(s)= (\phi_1(s), \phi_2(s)).$ We proceed under the following assumptions
on averages and correlations ($k=1,2):$

\begin{equation}
\label{BE0dj1}
E\phi_k(s)=0;
\end{equation}
\begin{equation}
\label{BE0dj}
E \phi_k(s_1) \overline{\phi_k(s_2)}= \sigma_k^2 \delta(s_1-s_2) \sqrt{\vert s_1 s_2\vert} + {\cal E}_0 \delta(s_1-s_2), {\cal E}_0 >0;  
\end{equation}
\begin{equation}
\label{BE0dj2}
E \phi_1(s_1) \overline{\phi_2(s_2)}= 2 \sqrt{{\cal E}_0} \sigma_{12} \delta(s_1-s_2) \vert s_1 s_2\vert^{1/4}, \; \sigma_{12} \in {\bf C};  
\end{equation}
\begin{equation}
\label{BE0dj3}
 \sigma_1^2= \sigma_2^2 = \vert \sigma_{12} \vert^2\equiv \sigma^2.
\end{equation}

{\bf Remark 4.} We stress the appearence of the additional term in (\ref{BE0dj}) comparing with (\ref{BE0d}). Physically this is the contribution (to correlations)
of the background field of the white noise type. We shall see that, in fact, ${\cal E}_0$ is the mean energy of this field. (It does not depend on $s).$
Its necessity was not evident in the case of the one-component signal (corresponding to a single quantum 
particle), so in section \ref{S1} we ignored the contribution of the background field.  However, in the case of bi-signals (corresponding to composite two particle systems) one cannot proceed classically without the background component. Surprisingly the presence of  
the background field started to play a role only in joint detection, or other way around: the presence of the background can be detected
only through joint measurement of correlated signals. We shall see that probabilities of joint detection predicted by QM (and tested
experimentally) correspond to well defined classical stochastic process only if the presence of the background field is taken into account.
This is a tricky situation. The contribuion of the background field is not directly present in quantum probabilities. It is eliminated
through calibration of detectors, see (\ref{BE0a0p10}). However, in the absence of this field ``prequantum stochastic process'' is not well defined.
(Of course, one may simply deny the existence of the prequantum classical process.)

\medskip

{\bf Remark 5.} (Dimension analysis) Here, cf. Remark 3, $\sigma_k^2   \sim {\it power}.$ The detection threshold 
${\cal E}_d \sim {\it energy}.$ From (\ref{BE0dj2}) we have that $E \phi_1(s_1) \overline{\phi_2(s_2)} = k \delta(s_1-s_2) \vert s_1 s_2\vert^{1/4},$
where $k^2 \times {\it time} \sim {\it energy}^2,$ i.e., $k^2 \sim \frac{{\it energy}^2}{{\it time}} \sim {\it energy} \times {\it power}.$ Hence, it is natural
to represent $k= k_0 \times \sigma_{12},$ where $\vert \sigma_{12}\vert^2 \sim {\it power}$ and  $ k_0^2\sim {\it energy}.$ We can select $k_0^2={\cal E}_0,$ the 
energy of vacuum fluctuations. The equality  (\ref{BE0dj3}) encodes matching of statistics of measurements on each of components $\phi_j(s), j=1,2,$ and  
joint measurement of these components. Hence, we consider very special class of signals.
 
\medskip

First, we show that this stochastic process is well defined.
Consider  the covariance function of this process
\begin{equation}
\label{BE0dj4}
D(s_1,s_2) = \left( \begin{array}{ll}
D_{11}(s_1, s_2)  & \; D_{12}(s_1, s_2) \\
 \; D_{21}(s_1, s_2) &  \;  D_{22}(s_1, s_2) \end{array}
 \right )
$$
$$
= \delta(s_1-s_2)
\left( \begin{array}{ll}
\sigma^2 \sqrt{\vert s_1 s_2\vert} + {\cal E}_0  & \; \;  2 \sqrt{{\cal E}_0} \sigma_{12}  \vert s_1 s_2\vert^{1/4} \\
 \; \;  2 \sqrt{{\cal E}_0} \bar{\sigma}_{12} \vert s_1 s_2\vert^{1/4} & \sigma^2 \sqrt{\vert s_1 s_2\vert} + {\cal E}_0 \end{array}
 \right ).
\end{equation}

\medskip

We now prove that the operator $\widehat{D}$ defined by the kernel (\ref{BE0dj4}) is positively defined. Take two $L_2$-functions, $y_1(s), y_2(s).$ We have
$$
\langle \widehat{D} y_1, y_2\rangle= \int (\sigma^2 \vert s\vert + {\cal E}_0) (\vert y_1(s)\vert^2 + \vert y_2(s)\vert^2) ds 
$$
$$
+  2 \sqrt{{\cal E}_0} \int \sqrt{\vert s\vert} (\sigma_{12} y_2(s)\bar{y_1}(s) + \bar{\sigma}_{12} y_1(s)\bar{y_2}(s)) ds = I_1 +I_2.
$$
We have 
$$
I_2 \geq - 4 \sqrt{{\cal E}_0} \vert \sigma_{12}\vert \int \vert y_1(s)\vert \vert y_2(s)\vert ds.
$$

Hence, $I_1+I_2 \geq$ 
$$
\int[(\sigma \sqrt{\vert s\vert}  \vert y_1(s) \vert
 -  \sqrt{{\cal E}_0} \vert y_2(s) \vert)^2 + 
(\sigma \sqrt{\vert s\vert}  \vert y_2(s) \vert -  \sqrt{{\cal E}_0} \vert y_2(s) \vert)^2  ds \geq 0.
$$

For each component of the bi-signal $\phi=(\phi_1, \phi_2),$ we consider the smoothed signal corresponding the integration window $\kappa,$
$\phi^\kappa=(\phi_1^\kappa, \phi_2^\kappa),$ see (\ref{BEb0}). Denote by ${\cal E}_k(s, \omega; \kappa)$ the energy 
of the $k$th component
of the $\kappa$-smooothed signal, i.e., ${\cal E}_k(s, \omega; \kappa)= \vert \phi_k^\kappa(s, \omega)\vert^2, k=1,2.$

In the absence of the background field,  we would have the threshold approaching detection conditions
\begin{equation}
\label{BE0a0}
{\cal E}_k(\tau_k, \omega; \kappa)=  {\cal E}_d, \; k=1,2,
\end{equation}
for each component, $\phi_k, k=1,2.$
(We assume that both detectors have the same detection threshold.) 

However, in the present model our signals are mixed with the background field. Denote the latter  by $\eta(s) \equiv \eta(s,\omega).$ Moreover, this field cannot be distilled from signals. 
There is no filter removing the background field. Its contribution can be strong enough to play an important role in production of clicks. We only can 
make cut-off in detectors by their calibration -- subtraction the energy of the background field. This field is very singular, so its energy 
for a fixed instance of time is not well defined. However, this problem is solved through using detectors with the integration windows given 
by functions of $g^{\kappa}$ type. They measure the energy of the smoothed $\eta:$
$$
\eta^\kappa (u) = \int \eta(s) g(u-s) ds= \langle g_u, \eta \rangle,
$$ 
where $g_u(s)=g(u-s).$
For such a detector, the energy contribution of the background field is given by
\begin{equation}
\label{BE0a00}
{\cal E}_0(u, \omega; \kappa)=  \vert \eta^\kappa (u) \vert^2= \vert \langle g_u, \eta \rangle\vert^2.
\end{equation}
Hence, the detection condition for each component of the bi-signal can be modified from (\ref{BE0a0}) to 
\begin{equation}
\label{BE0a0X}
{\cal E}_k(\tau_k, \omega; \kappa) - {\cal E}_0(\tau_k, \omega; \kappa)=  {\cal E}_d
\end{equation}
or
\begin{equation}
\label{BE0a0p}
{\cal E}_k(\tau_k, \omega; \kappa)=   {\cal E}_d^\prime,
\end{equation}
where ${\cal E}_d^\prime=  {\cal E}_d + {\cal E}_0(\tau_k, \omega; \kappa) $ is the calibrated threshold. 
However, the threshold ${\cal E}_d^\prime$ is random. So, it is unuseful for the practical purpose. 
The (random) contribution of the background is unknown. Therefore in practice the detection condition (\ref{BE0a0p})
is changed to coarser condition with {\it calibration by the mean value of the detected energy of the background field.}

First we find  this mean value for the fixed (i.e., nonrandom) $\tau.$
We use the general result on quadratic forms of Gaussian random variables valued in Hilbert spaces \cite{KH1}, see equality (\ref{BL0}) 
in appendix.
Consider in $L_2$ the operator $\widehat{A} \equiv \widehat{A}_{\tau;\kappa}=\vert g_\tau\rangle \langle g_\tau\vert,$ where, as always,  $g_\tau(s) = g(\tau-s)$ 
and the function $g$ was defined in (\ref{KAPPA}). Set $f_A(y) = \langle \widehat{A} y, y\rangle, y \in L_2,$ the quadratic form corresponding 
to the operator $\widehat{A}.$ By (\ref{BL0}) we obtain
$$
E {\cal E}_0(\tau, \omega; \kappa) = E f_A(\eta^{\kappa})= {\cal E}_0 \rm{Tr} \widehat{A}= {\cal E}_0 \Vert g_\tau\Vert= {\cal E}_0.
$$
This quantity does not depend on $\tau$ and this is not surprising, since the background field is translation invariant. If $\tau$ is random (as it is in (\ref{BE0a00})), then we can use the formula of total probability:
$$
E {\cal E}_0(\tau(\omega), \omega; \kappa)= \int_0^\infty E[ {\cal E}_0(\tau(\omega), \omega; \kappa)\vert \tau(\omega)=\tau] P(\tau(\omega)=\tau) d\tau=
{\cal E}_0.
$$
Now we modify the detection condition (\ref{BE0a0p}) and proceed with conditions ($k=1,2)$
\begin{equation}
\label{BE0a0p10}
 {\cal E}_k(\tau_k, \omega; \kappa) - {\cal E}_0 = {\cal E}_d 
 \end{equation}
 or 
\begin{equation}
\label{BE0a0p1}
 {\cal E}_k(\tau_k, \omega; \kappa)= {\cal E}_d^\prime 
 \end{equation}
  where 
\begin{equation}
\label{BE0a0p10h}
{\cal E}_d^\prime= {\cal E}_0 + {\cal E}_d.
\end{equation}

 For each component of the bi-signal, we repeat the scheme of sections \ref{S1},  \ref{S2}, but with the new threshold given by 
(\ref{BE0a0p10h}).  

The only difference is that the process $\phi_k(s)$ has the covariance operator 
$\widehat{D}_{kk}=  \widehat{D}^{(0)}_{kk} + {\cal E}_0 I,$
where $\widehat{D}^{(0)}_{kk}$ is the covariance operator of the process which was considered in section \ref{S1}.  We have, see appendix, 
\begin{equation}
\label{BE0iiiK}
E  {\cal E}_k(\tau, \omega; \kappa) = \rm{Tr} \widehat{D}_{kk} \widehat{A}=  \rm{Tr} \widehat{D}^{(0)}_{kk} \widehat{A} +  {\cal E}_0 \rm{Tr} \widehat{A}=
 \rm{Tr} \widehat{D}^{(0)}_{kk} \widehat{A} +  {\cal E}_0.
\end{equation}
Therefore the detection condition (\ref{BE0a0p1}) with (\ref{BE0a0p10h}) implies (after averaging and the use of the formula of total probability) the same considtion
as in section \ref{S1}
\begin{equation}
\label{BE0iii}
 \rm{Tr} \widehat{D}^{(0)}_{kk} \widehat{A}= \bar{\tau} \sigma^2 (1+ O(\kappa/\bar{\tau})) = {\cal E}_d.
\end{equation}
Thus the contribution of the background field was completely excluded -- through the proper calibration of detectors. (We stress again
that this can be done only``afterward'', i.e., on the level of detectors and not fields; this is a crucial point of our approach to QM, as theory of measurements
with threshold's type and properly calibrated detectors.) 

Now we consider the joint clicks in detectors corresponding to the components of the bi-signal.  Thus
(\ref{BE0a0p}) holds for both $k$s and moreover the instances of detection for corresponding 
detectors, $\tau_k= \tau_k(\omega),$ are constrained by the equality:
\begin{equation}
\label{BE0a01} 
\tau=\tau_1= \tau_2.
\end{equation}

{\bf Remark 6.} Of course, in the real experiment we cannot proceed with the precise coincidance of instances of detection. One has to use the 
joint detection time window, say $v,$ and proceed under the condition
\begin{equation}
\label{BE0a02} 
\vert \tau_1 - \tau_2 \vert \leq v.
\end{equation}
In our indeal model we ignore this experimental technicality. Opposite to the model from \cite{Raedt}, the presence
the joint detection time window $v\not=0$  in real experiments does not play a crucial role in our model, i.e., we can obtain quantum correlations
even for $v=0.$

\medskip

We now find average of the joint detection time $\tau.$   
The system of equalities (\ref{BE0a0p10}), $k=1,2,$ and (\ref{BE0a01}) imply
\begin{equation}
\label{BE0a0y}
({\cal E}_1(\tau(\omega), \omega; \kappa) - {\cal E}_0) 
( {\cal E}_2(\tau(\omega), \omega; \kappa) - {\cal E}_0)  =  {\cal E}_d^2.
\end{equation}
We take the average of the both sides
\begin{equation}
\label{BE0a0y1}
E ({\cal E}_1(\tau(\omega), \omega; \kappa) - {\cal E}_0)  
({\cal E}_2(\tau(\omega), \omega; \kappa) - {\cal E}_0)  =  {\cal E}_d^2.
\end{equation}
or 
\begin{equation}
\label{BE0a0y1X}
E {\cal E}_1 (\tau(\omega), \omega; \kappa ) {\cal E}_2 (\tau(\omega), \omega; \kappa ) - {\cal E}_0 (E  {\cal E}_1 (\tau(\omega), \omega; \kappa )  +
 E {\cal E}_2 (\tau(\omega), \omega; \kappa)) +   
{\cal E}_0^2  =  {\cal E}_d^2.
\end{equation}
We start with the first term in the left-hand side of this equality.
We shall again use the formula of total probability
$$
E {\cal E}_1 (\tau(\omega), \omega; \kappa) {\cal E}_2 (\tau(\omega), \omega; \kappa) 
$$
\begin{equation}
\label{TPROB}
 = \int_0^\infty E 
[{\cal E}_1 (\tau(\omega), \omega; \kappa ) {\cal E}_2 ( \tau(\omega), \omega; \kappa ) 
\vert \tau(\omega)= \tau ] P ( \tau(\omega)= \tau ) d\tau.
\end{equation}
For the fixed $\tau,$ we have to find the correlation of two quadratic forms of the component of the Gaussian bi-signal satisfying the aforementioned assumptions.
We again use the general result on quadratic forms of Gaussian random variables valued in Hilbert spaces \cite{KH1}, see equation  (\ref{BL}) in appendix.
Consider again the operator $\widehat{A}=\vert g_\tau\rangle \langle g_\tau\vert.$ and its quadratic form 
$f_A(y).$ Then by (\ref{BL}) we have
\begin{equation}
\label{BE0dj4kp}
E {\cal E}_1(\tau, \omega; \kappa) {\cal E}_2(\tau, \omega; \kappa)= E f_A(\phi_1) f_A(\phi_2)
\end{equation}
$$
= \rm{Tr}   \widehat{D}_{11} \widehat{A} \; \rm{Tr}    \widehat{D}_{22} \widehat{A}  + \langle  \widehat{A} \otimes  \widehat{A} D_{12}, D_{12}\rangle =J_1 + J_2,
$$
where 
\begin{equation}
\label{BE0dj4k}
\widehat{D} = \left( \begin{array}{ll}
\widehat{D}_{11}  & \; \widehat{D}_{12} \\
 \; \widehat{D}_{21} & \widehat{D}_{22} \end{array}
 \right )
\end{equation}
is the covariance operator corresponding to the kernel $D(s_1,s_2).$
We start with the last term. It is determined by the off-diagonal term $D_{12}(s_1,s_2)$ of the covariance function $D(s_1, s_2):$
$$
J_2= \Big\vert \int \int g_\tau(s_1) g_\tau(s_2) D_{12}(s_1, s_2) ds_1 ds_2\Big\vert^2=
\Big\vert \frac{2 \sqrt{{\cal E}_0} \sigma_{12}}{\kappa}\int_{\tau-\kappa}^\tau \sqrt{\vert s\vert} ds\Big\vert^2
$$
$$
=
\Big\vert \frac{4\sqrt{{\cal E}_0} \sigma_{12}}{3\kappa} [\tau^{3/2} -(\tau - \kappa)^{3/2}]\Big\vert^2=
4 {\cal E}_0 \sigma^2 \tau (1+ O(\kappa/\tau))^2 .
$$
$$
=4 {\cal E}_0 \sigma^2 \tau (1+ O(\kappa/\tau)) , \;     \kappa/\tau \to 0.
$$
Now we consider 
$$
J_1= (\rm{Tr}   \widehat{D}_{11} \widehat{A})(\rm{Tr}   \widehat{D}_{22} \widehat{A}) =(\rm{Tr}   \widehat{D}_{11} \widehat{A})^2= \langle \widehat{D}_{11} g_\tau, g_\tau\rangle^2.
$$
We have 
$$
\langle \widehat{D}_{11} g_\tau, g_\tau\rangle =  \int (\sigma^2 \vert s_1 \vert + {\cal E}_0) g_\tau^2(s) ds=
\frac{1}{\kappa} \int_{\tau-\kappa}^\tau (\sigma^2 s + {\cal E}_0) ds= \sigma^2 \tau ( 1+ O(\kappa/\tau)) + 
{\cal E}_0.
$$
Hence,
$$
J_1= (\sigma^4 \tau^2 + 2 \sigma^2 \tau {\cal E}_0 + {\cal E}_0^2) (1+O(\kappa/\tau)). 
$$
and 
$$
E {\cal E}_1(\tau, \omega; \kappa) {\cal E}_2(\tau, \omega; \kappa)= [ 4 {\cal E}_0 \sigma^2 \tau + (\sigma^4 \tau^2 + 2 \sigma^2 \tau {\cal E}_0 + {\cal E}_0^2)] (1+O(\kappa/\tau))
$$
$$
 \approx 4 {\cal E}_0 \sigma^2 \tau + (\sigma^4 \tau^2  +2\sigma^2\tau {\cal E}_0 + {\cal E}_0^2). 
\; \kappa/\tau \to 0. 
$$ 
We now turn to the formula of total probability (\ref{TPROB}) and we obtain
\begin{equation}
\label{TPROBX}
E {\cal E}_1(\tau(\omega), \omega; \kappa) {\cal E}_2(\tau(\omega), \omega; \kappa)  
\approx  \int_0^\infty [ \sigma^2 \tau + (\sigma^4 \tau^2  +2\sigma^2\tau {\cal E}_0 + {\cal E}_0^2)] P(\tau(\omega)= \tau ) d\tau
$$
$$
=  4 {\cal E}_0  \sigma^2 \bar{\tau} + ( \sigma^4 \bar{\tau^2} + 2 \sigma^2 \bar{\tau} {\cal E}_0 + {\cal E}_0^2), \; \kappa/\tau \to 0.
\end{equation}
Finally, turn to the basic detection condition (\ref{BE0a0y1X}):
$$
4 {\cal E}_0 \sigma^2 \bar{\tau} + ( \sigma^4 \bar{\tau^2} + 2 \sigma^2 \bar{\tau} {\cal E}_0 + {\cal E}_0^2) -
2 {\cal E}_0 (\sigma^2 \bar{\tau} +  {\cal E}_0) + {\cal E}_0^2 \approx   {\cal E}_d^2.
$$
or 
\begin{equation}
\label{TPpi}
4 {\cal E}_0 \sigma^2 \bar{\tau} + \sigma^4 \bar{\tau^2} \approx   {\cal E}_d^2.
\end{equation}
Suppose now that 
\begin{equation}
\label{TPROB1}
\sigma^4 \bar{\tau^2}<<  {\cal E}_0 \sigma^2 \bar{\tau} 
\end{equation}
Thus the second term in the left-hand side of the equality (\ref{TPpi}) is essentially less than the second term. Hence, we have
 \begin{equation}
\label{TPpiX}
4 {\cal E}_0 \sigma^2 \bar{\tau} \approx   {\cal E}_d^2.
\end{equation}
Now we analyze the condition (\ref{TPROB1}). It can be written as
\begin{equation}
\label{TPpkk}
\sigma^2 << \frac{{\cal E}_0 \bar{\tau}}{\bar{\tau^2}}
\end{equation}
or
\begin{equation}
\label{TPpkk1}
\sigma^2 << \frac{{\cal E}_0}{\bar{\tau}} \frac{\bar{\tau}^2}{\bar{\tau^2}}.
\end{equation}
By the Cauchy-Bunyakovsky inequality $\bar{\tau}^2\leq \bar{\tau^2}.$ Hence, we have
\begin{equation}
\label{TPpkk2}
\sigma^2 << \frac{{\cal E}_0}{\bar{\tau}}.
\end{equation}
We state again that the quantity $\sigma^2$ has the dimension of {\it signal's power.} Hence, the condition (\ref{TPpkk}) is a constraint to signal's power. The quantity $\frac{{\cal E}_0}{\bar{\tau}}$ is average power of the background field (vacuum flcutuations) during the period of 
detection (``click's production''). Hence, our approach is about detection of {\it weak signals on the strong random background.}

\section{Probability of coincidence of clicks}
\label{S2}

Hence, during a long period of time $T$ a pair of detectors clicks  jointly $N_{\rm{click}}$-times, where
\begin{equation}
\label{BE0c3T} 
 N_{\rm{click}} \approx \frac{T}{\bar{\tau}}  \approx  \frac{4 {\cal E}_0 \sigma^2 T}{ {\cal E}_d^2},
\end{equation}
where $\kappa/\bar{\tau} \to 0$ and the condition (\ref{TPpkk2}) holds. 
To find the probability of detection and match the real detection scheme which is used quantum experiments \cite{AS}, we have to use a proper normalization.
This is again an important point of our considerations, cf. section \ref{S2}. In QM-experiments with composite systems probabilities are obtained
through normalization corresponding to the sum of joint clicks in all pairs of detectors involved in the experiment.
For example, for measurement of spin projections for a pair of electrons (e.g., entangled) to some axes $a$ and $b,$ we use 
two pairs of detectors: $D_{1+},D_{1-},$ spin up and spin down for the first electron, and $D_{2+},D_{2-},$ spin up and spin down for the second electron. We collect the numbers of clicks for the pairs of detectors:  $N_{\rm{click}}(++)$ for $D_{1+},D_{2+},$ ..., 
$N_{\rm{click}}(--)$ for $D_{1-},D_{2-}.$ Then we compute the total sum of clicks $N=  N_{\rm{click}}(++) + 
N_{\rm{click}}(+-) + N_{\rm{click}}(-+) + N_{\rm{click}}(--)$ and it is used as the normalization factor for computing of probabilities, e.g., $P(++)= \frac{N_{\rm{click}}(++)}{N}.$ We repeat this scheme in the general case of detection of observable with discrete spectrum.
Suppose that each component of a random bi-signal $\phi(s)=(\phi_1(s), \phi_2(s))$ is complex vector 
$\phi_1(s)= (\phi_1(i, s)))_{i=1}^m,  \phi_2(s)= (\phi_2(i, s)))_{i=1}^m.$ 
Consider a {\it Gaussian} bi-signal. Assumptions (\ref{BE0dj1})--(\ref{BE0dj3}) are modified ($i, j=1,...,m):$
\begin{equation}
\label{BE0dj1a}
E\phi_k(i, s)=0;
\end{equation}
\begin{equation}
\label{BE0dja}
E \phi_k(i, s_1) \overline{\phi_k(j, s_2)}= \sigma_k^2(ij) \delta(s_1-s_2) \sqrt{\vert s_1 s_2\vert} + {\cal E}_0 \delta(s_1-s_2), {\cal E}_0 >0;  
\end{equation}
\begin{equation}
\label{BE0dj2a}
E \phi_1(i, s_1) \overline{\phi_2(j, s_2)}= 2 \sqrt{{\cal E}_0} \sigma_{12}(ij) \delta(s_1-s_2) \vert s_1 s_2\vert^{1/4}, \; 
\sigma_{12}(ij) \in {\bf C};  
\end{equation}
To match completely the QM-theory, the condition (\ref{BE0dj3}), coupling between powers of signal's components $\sigma_k^2$ 
and ``power of correlations between component'' $\sigma_{12},$  has to be generalized to the case of vector processes in rather 
tricky way, see (\ref{BE0dj3b}). To clarify the main points of derivation of probabilities
for coincidences, we start with a simpler stochastic model which will reproduce probabilities for coincidences, but not yet probabilities
for measurements on each fixed component.\footnote{Of course, by knowing probabilities for coincidences we can derive probabilities
for measurements on signals $\phi_k, k=1,2,$ by using the laws of (classical) probability theory. However, we want to introduce such a random bi-signal $\phi=(\phi_1, \phi_2)$ that by performing measurement (with detectors of the thershold type and the background field calibration) only on $\phi_k$ we shall obtain the corresponding quantum probability.}  In this section we proceed with stochastic processes
satisfying conditions (\ref{BE0dj1a})--(\ref{BE0dj2a}) and
\begin{equation}
\label{BE0dj3a}
 \sigma_1^2(ij)= \sigma_2^2(ij) = \vert \sigma_{12}(ij) \vert^2\equiv \sigma^2 (ij).
\end{equation}
We have 
\begin{equation}
\label{BE0co} 
 N_{\rm{click}}(ij) \approx \frac{T}{\bar{\tau}_{ij}}  \approx  \frac{4 {\cal E}_0 \sigma^2(ij) T}{ {\cal E}_d^2}.
\end{equation}
The total number of clicks
\begin{equation}
\label{BE0co1} 
 N_{\rm{click}}  = \sum_{ij} N_{\rm{click}}(ij) \approx \frac{4 {\cal E}_0 \Sigma^2 T}{ {\cal E}_d^2},
\end{equation}
where
$\Sigma^2= \sum_{ij} \sigma^2(ij)$ is the total averaged power of the bi-signal. The probability of detection for the pair of
detectors $D_{1i}$ and $D_{2j}$ is given by 
\begin{equation}
\label{BE0co2} 
P_{\rm{click}}(ij)= \frac{N_{\rm{click}}(ij)}{N_{\rm{click}}}\approx \frac{\sigma^2(ij)}{\Sigma^2}.
\end{equation}
In fact, this is the Born's rule. Consider the complex vector 
\begin{equation}
\label{psi} 
\psi= (\sigma_{12}(ij)).\end{equation} (Its dimension is $m^2,$ i.e., the squared 
dimension of the state space of  components $\phi_k.)$  We remark that it is not normalized by one. Its squared norm is $\Vert \psi \Vert^2 = \Sigma^2.$ We normalize this vector: 
\begin{equation}
\label{Psi} 
\Psi= \frac{\psi}{\Vert \psi\Vert}.
\end{equation}  By our interpretation of 
QM this is a state vector. (So, the quantum state vector of a composite system is constructed from correlations between components of the 
``prequantum stochastic process''; the quantum system is its symbolic representation in the operational formalism called QM.) In such 
notation we have
\begin{equation}
\label{BE0co2X} 
P_{\rm{click}}(ij)= \vert \Psi(ij)\vert^2.
\end{equation}

In our approach the QM-formalism is the operational formalism in which connection of the quantum state vector with correlations 
inside ``prequantum random signals'' is ignored. In QM the $\Psi$-state is invented formally; then it is used to find correlations.
In our approach the $\Psi$-state is nothing else than the symbolic representation of correlations in the classical signals.

\section{The final stochastic model}

We now consider a more tricky (classical) stochastic process. It satisfies the conditions (\ref{BE0dj1a})--(\ref{BE0dj2a}) and, instead 
of condition (\ref{BE0dj3a}), the condition: 
\begin{equation}
\label{BE0dj3b}
 \sigma_1^2(ij)=  \sum_n \sigma_{12}(in) \bar{\sigma}_{12}(jn), \;  
\sigma_2^2(ij)= \sum_n \sigma_{12}(ni) \bar{\sigma}_{12}(nj).
\end{equation}
Later we shall write this condition in the matrix form, by using the matrix of cross-correlations 
$\hat{\sigma}_{12}=(\sigma_{12}(ij)).$

First we show that this process also reproduces the quantum probability for coincidence measurements on components $\phi_1$ and 
$\phi_2,$ cf. section \ref{S2}. We slightly modify the results of calculation in section \ref{SA}. 
We set ${\cal E}_k(i, s, \omega; \kappa)= \vert \phi_k^\kappa(i,s, \omega)\vert^2, k=1,2; i=1,...,m.$
Generalizing (\ref{BE0iiiK}) and (\ref{BE0iii}), we obtain 
\begin{equation}
\label{BE0iiiKw}
E  {\cal E}_k(i, \tau, \omega; \kappa)=\tau \sigma_k^2(ii) (1+ O(\kappa/\tau )) +  {\cal E}_0, k=1,2.
\end{equation}

We now find  $E {\cal E}_1(i, \tau, \omega; \kappa) {\cal E}_2(j, \tau, \omega; \kappa)= J_1(ij) + J_2(ij),$ cf. (\ref{BE0dj4kp}):
$$
J_2(ij) = 4 {\cal E}_0 \vert \sigma_{12}(ij) \vert^2 \tau (1+ O(\kappa/\tau)) , \;     \kappa/\tau \to 0;
$$
$$
J_1(ij)= (\sigma_1^2(ii) \tau + {\cal E}_0)(\sigma_2^2(jj) \tau + {\cal E}_0) (1+ O(\kappa/\tau))
$$
$$
= 
[\sigma_1^2(ii)\sigma_2^2(jj) \tau^2 + {\cal E}_0 \tau 
(\sigma_1^2(ii) + \sigma_2^2(ii)) + {\cal E}_0^2] (1+ O(\kappa/\tau)),  \kappa/\tau \to 0.
$$
As always by using the formula of total probability,  we obtain
$$
E {\cal E}_1(i, \tau(\omega), \omega; \kappa) {\cal E}_2(j, \tau(\omega), \omega; \kappa) \approx 
$$
\begin{equation}
\label{BE0iiiKa}
4 {\cal E}_0 \vert \sigma_{12}(ij)\vert^2 \bar{\tau} +\sigma_1^2(ii)\sigma_2^2(jj) \bar{\tau^2} + {\cal E}_0 \bar{\tau} 
(\sigma_1^2(ii) + \sigma_2^2(ii)) + {\cal E}_0^2, \kappa/\tau \to 0. 
\end{equation}
Bu using the condition (\ref{BE0a0y1X}) for $i$th and $j$th coordinates of the signals $\phi_1$ and $\phi_2$ we obtain
\begin{equation}
\label{BE0iiiKa1}
4 {\cal E}_0 \vert \sigma_{12}(ij)\vert^2 \bar{\tau} +\sigma_1^2(ii)\sigma_2^2(jj) \bar{\tau^2} \approx {\cal E}_d^2. 
\end{equation}
Under the assumption
\begin{equation}
\label{BE0iiiKab}
\sigma_1^2(ii)\sigma_2^2(jj) \bar{\tau^2} << {\cal E}_0 \vert \sigma_{12}(ij)\vert^2 \bar{\tau},  
\end{equation}
we obtain the detection condition
\begin{equation}
\label{BE0iiiKa1X}
4 {\cal E}_0 \vert \sigma_{12}(ij)\vert^2 \bar{\tau}  \approx {\cal E}_d^2. 
\end{equation}
which is the basic to derive detection probabilities for coincidence of clicks.

The condition (\ref{BE0iiiKab}) implies that 
\begin{equation}
\label{BE0iiiKabX}
\sum_{ij} \sigma_1^2(ii)\sigma_2^2(jj) << \frac{{\cal E}_0}{\tau}\sum_{ij}  \vert \sigma_{12}(ij)\vert^2, 
\end{equation}
i.e., for 
\begin{equation}
\label{TTT}
\sigma_k^2= \sum_{i} \sigma_k^2(ii), k=1,2, \vert\sigma_{12}\vert^2=  \sum_{ij}  \vert \sigma_{12}(ij)\vert^2,
\end{equation}
we have
\begin{equation}
\label{BE0iiiKabc}
\frac{\sigma_1^2 \sigma_2^2}{\vert\sigma_{12}\vert^2}
 << \frac{{\cal E}_0}{\tau}.
\end{equation}
Quantities $\sigma_k^2, k=1,2,$ have the meaning of average powers of signal's components $\phi_k;$ the physical meaning of the
quantity  $\vert\sigma_{12}\vert^2$ is not straightforward. Formally, it can be considered as 
``power of correlations between components''. By using this terminology we can say that our (coming) derivation of Bornn's rule 
is valid for signals of sufficiently low relative power (comparing with power of the background field) of signal's components 
comparing with power of correlations between components.
nal
Consider again the complex vector $\psi =(\sigma_{12}(ij)),$ see (\ref{psi}), and its normalization $\Psi,$ see (\ref{Psi}).
 Starting with detection condition (\ref{BE0iiiKa1}) and repeating the steps of the derivation of section \ref{S2}, we obtain
 again Born's rule for detection of coincidences. Now we show that even for each sigle detector we obtain the quantum formula for probability.
 
 By using the formula of total probability we obtain from (\ref{BE0iiiKw})
$E  {\cal E}_k(i, \tau, \omega; \kappa) \approx \bar{\tau} \sigma_k^2(ii) +  {\cal E}_0, k=1,2.$
For the $i$th coordinate of the component $\phi_k$ we have the detection condition 
$ {\cal E}_k(i, \tau_k(i), \omega; \kappa)= {\cal E}_d^\prime $
where ${\cal E}_d^\prime= {\cal E}_0 + {\cal E}_d.$
Hence, $\bar{\tau} \sigma_k^2(ii)= {\cal E}_d.$
The number of clicks is given by $N_{\rm{click}, k}(i)= \frac{T \sigma_k^2(ii)}{{\cal E}_d};$ the total number of clicks 
at all detectors for coordinates $\phi_k(i)$ of  the component $\phi_k$ is given by
$N_{\rm{click}, k}= \sum_i N_{\rm{click}, k}(i)= \frac{T \sigma_k^2}{{\cal E}_d},$ see (\ref{TTT}). Hence,
$P_{\rm{click},k}(i) = \frac{\sigma_k^2(ii)}{\sigma_k^2}.$

Now, for the vector $\Psi$ consider the corresponding projection operator $\rho_\Psi= \vert \Psi \rangle \langle \Psi \vert$ 
and its partial traces $\rho_\Psi^{(k)}, k=1,2.$ We also introduce operators $\hat{\sigma}_k^2=(\sigma_k^2(ij)).$ We have
$\rm{Tr} \hat{\sigma}_k^2= \sigma_k^2$ and the equality (\ref{BE0dj3b}) implies that $\rho_\Psi^{(k)}=   \hat{\sigma}_k^2/ \rm{Tr} \hat{\sigma}_k^2.$ The final formula derived for the detection probability has the form
$$ 
P_{\rm{click},k}(i) = \rm{Tr} \rho_\Psi^{(k)} \widehat{C}_j,
$$ 
where $\widehat{C}_j = \vert e_j \rangle \langle e_j \vert$ is the projector onto the vector $e_j$ corresponding to the detection
in the $i$th channel for the $\phi_k.$ 

We state again that each measurement under consideration corresponds to expansion of the signal's components with respect 
to some bases, say $\{e_{ki}\}, k=1,2,$ in the state spaces of signal's components $\phi_k.$ Detectors measure signals
$\phi_k(i)= \langle \phi_k, e_{ki}\rangle, i=1,...,m.$ 

\section{Violation of CHSH inequality} 

We borrow from QM the singlet state $\Psi= \frac{1}{\sqrt{2}}
(\vert +\rangle \vert -\rangle - \vert -\rangle \vert +\rangle),$ where  $e_{\pm}= \vert \pm\rangle$ is 
$z$-polarziations basis. The simplest way to select  the proper classical correlations is to  identify
$\psi,$  see (\ref{psi}), with $\Psi:$   $\sigma(12)= - \sigma(21)=   \frac{1}{\sqrt{2}}.$ These correlations detremine
the classical random bi-signal $\phi(s)= (\phi_1(s), \phi_2(s)).$ Each component is valued in the two dimensional complex space:
$\phi_j(s)= \phi_j(+, s) e_+ + \phi_j(-, s) e_-, j=1,2.$ We fix two angles $\theta_1, \theta_2$ and the  corresponding bases: $e_{\pm}^{\theta_j}.$ Consider expansions of the bi-signal's components: $\phi_j(s)= \phi_{\theta_j}(+, s) e_+^{\theta_j} + \phi_{\theta_j}(-, s) e_-^{\theta_j}.$
Consider probabilities for joint measurements of the signals $\phi_{\theta_1}(\pm, s)$ and $\phi_{\theta_2}(\pm, s).$ Since they 
coincide with the corresponding quantum probabilities,
these probabilities  for the joint detection of  
classical random singals by the threshold type and properly calibrated detectors violate CHSH inequality.

The QM state $\Psi$ determines correlations $\sigma_{12}(ij)$ up to a normalization factor. This 
state corresponds to a family of classical random fields. So, the correspondence between classical and quantum models
is not one-to-one.

\section{Appendix: Gaussian integrals}

Let $W$ be a real Hilbert space.
Consider  a $\sigma$-additive Gaussian measure $p$ on the
$\sigma$-field of Borel subsets of $W.$ This measure is determined
by its covariance operator $B:  W \to W$ and mean value $m \in W.$
For example, $B$ and $m$ determine the Fourier transform of $p:$
$$
\tilde p(y)= \int_W e^{i(y, \phi)} dp (\phi)=
e^{\frac{1}{2}(By, y) + i(m, y)}, y \in W.
$$
(In probability theory it is called the characteristic functional of the probability distribution $p.)$
In what follows we restrict our considerations to {\it Gaussian
measures with zero mean value}: $ (m,y) = \int_W(y, \psi) d p
(\psi)= 0 $ for any $y \in W.$ Sometimes there will be used the
symbol $p_B$ to denote the Gaussian measure with the covariance
operator $B$ and $m=0.$ We recall that the covariance operator $B$
is defined by its bilinear form
$ (By_1, y_2)=\int (y_1, \phi) (y_2, \phi) dp(\phi),
y_1, y_2 \in W$

Let $Q$ and $P$ be two copies of a real Hilbert space. Let us
consider their Cartesian product $H=Q \times P,$  ``phase space,''
endowed with the symplectic operator $J= \left( \begin{array}{ll}
 0&1\\
-1&0
\end{array}
 \right ).$
Consider the class of Gaussian measures (with zero mean value)
which are invariant with respect to the action of the operator
$J;$ denote this class ${\cal S}(H).$ It is easy to show that $p \in
{\cal S}(H)$ if and only if its covariance operator commutes with the
symplectic operator, \cite{KH1}.
 
As always, we consider complexification of $H$ (which will be
denoted by the same symbol), $H=Q\oplus i P.$ The complex scalar
product is denoted by the symbol $\langle \cdot, \cdot \rangle.$
The space of bounded Hermitian operators acting in $H$ is denoted by
the symbol ${\cal L}_s (H).$

We introduce the complex covariance operator of a measure $p$ on
the complex Hilbert space $H:\;$
$
\langle Dy_1, y_2 \rangle = \int_H \langle y_1, \phi \rangle \langle \phi, y_2 \rangle d p (\phi).
$
Let $p$ be a measure on the Cartesian product $H_1 \times H_2$ of
two complex  Hilbert spaces. Then its covariance operator has the block
structure
\begin{equation}
\label{BLOCK}
D = \left( \begin{array}{ll}
 D_{11} & D_{12}\\
D_{21} & D_{22}\\
 \end{array}
 \right ),
 \end{equation}
where $D_{ii} : H_i \to H_i$ and $D_{ij}: H_j \to H_i.$ The
operator is Hermitian. Hence $D_{ii}^* = D_{ii},$ and $D_{12}^*
= D_{21}.$

Let $H$ be a complex Hilbert space and let $\widehat A \in {\cal
L}_s (H).$ We consider its quadratic form (which will play an important role
in our further considerations)
$
\phi \to f_A(\phi) = \langle \widehat{A} \phi,\phi\rangle.
$
We make a trivial, but ideologically important remark:  $f_A: H \to {\bf R} ,$ is a ``usual
 function'' which is defined point wise.  We use the equality, see, e.g., \cite{KH1}:
\begin{equation}
\label{BL0} \int_H f_A(\phi) dp_D (\phi)={\rm Tr}\; D \widehat A
\end{equation}

\medskip

Let $p$ be a Gaussian measure of the class ${\cal S}(H_1 \times H_2)$ with the
(complex) covariance operator $D$ and let operators $\widehat A_i$ belong to the class ${\cal
L}_s (H_i), i= 1,2.$ Then
\begin{equation}
\label{BL}
\int_{H_1 \times H_2}
f_{A_1}( \phi_1) f_{A_2}(\phi_2) d p (\phi) =
{\rm Tr} D_{11} \widehat A_1 \; {\rm Tr}D_{22} \widehat A_2 +
{\rm Tr}D_{12} \widehat A_2 D_{21} \widehat A_1
\end{equation}

\medskip

This equality   is a consequence of the following general result \cite{KH1}:

 Let $p \in {\cal S}(H)$ with the
(complex) covariance operator $D$ and let $\widehat A_i \in {\cal L}_s (H).$ Then
\begin{equation}
\label{YY1} \int_{H} f_{A_1}( \phi) f_{A_2}(\phi)   dp (\phi), = {\rm Tr} D
\widehat A_1 {\rm Tr} D \widehat A_2 + {\rm Tr} D \widehat A_2 D
\widehat A_1.
\end{equation}

\end{document}